\documentclass[twocolumn]{aastex7}

\usepackage{graphicx}
\usepackage{amsmath}
\usepackage{hyperref}
\usepackage{longtable}
\usepackage{booktabs}

\newcommand{\te}{t_{\rm E}}
\newcommand{\thetae}{\theta_{\rm E}}

\newcommand{\pie}{\pi_{\rm E}}
\newcommand{\pirel}{\pi_{\rm rel}}

\newcommand{\dl}{D_{\rm L}}

\def\e{{\rm E}}

\begin{document}

\title{Three binary-source binary-lens microlensing events from the 2024 microlensing
campaign }
\shorttitle{Three binary-source binary-lens microlensing events}


\author{Cheongho Han}
\affiliation{Department of Physics, Chungbuk National University, Cheongju 28644, Republic of Korea}
\email{cheongho@astroph.chungbuk.ac.kr}
\author{Chung-Uk Lee}
\affiliation{Korea Astronomy and Space Science Institute, Daejon 34055, Republic of Korea, \thanks{Corresponding author: \texttt{leecu@kasi.re.kr}}}
\email{leecu@kasi.re.kr}
\author{Andrzej Udalski} 
\affiliation{Astronomical Observatory, University of Warsaw, Al.~Ujazdowskie 4, 00-478 Warszawa, Poland}
\email{udalski@astrouw.edu.pl} 
\collaboration{5}{(Leading authors)}
\author{Michael D. Albrow}   
\affiliation{University of Canterbury, Department of Physics and Astronomy, Private Bag 4800, Christchurch 8020, New Zealand}
\email{michael.albrow@canterbury.ac.nz}
\author{Sun-Ju Chung}
\affiliation{Korea Astronomy and Space Science Institute, Daejon 34055, Republic of Korea}
\email{sjchung@kasi.re.kr}
\author{Andrew Gould}
\affiliation{Department of Astronomy, Ohio State University, 140 West 18th Ave., Columbus, OH 43210, USA}
\email{gould.34@osu.edu}
\author{Youn Kil Jung}
\affiliation{Korea Astronomy and Space Science Institute, Daejon 34055, Republic of Korea}
\affiliation{University of Science and Technology, Daejeon 34113, Republic of Korea}
\email{younkil21@gmail.com}
\author{Kyu-Ha~Hwang}
\affiliation{Korea Astronomy and Space Science Institute, Daejon 34055, Republic of Korea}
\email{kyuha@kasi.re.kr}
\author{Yoon-Hyun Ryu}
\affiliation{Korea Astronomy and Space Science Institute, Daejon 34055, Republic of Korea}
\email{yhryu@kasi.re.kr}
\author{Yossi Shvartzvald}
\affiliation{Department of Particle Physics and Astrophysics, Weizmann Institute of Science, Rehovot 76100, Israel}
\email{yossishv@gmail.com}
\author{In-Gu Shin}
\affiliation{Department of Astronomy, Westlake University, Hangzhou 310030, Zhejiang Province, China}
\email{ingushin@gmail.com}
\author{Jennifer C. Yee}
\affiliation{Center for Astrophysics $|$ Harvard \& Smithsonian 60 Garden St., Cambridge, MA 02138, USA}
\email{jyee@cfa.harvard.edu}
\author{Weicheng Zang}
\affiliation{Center for Astrophysics $|$ Harvard \& Smithsonian 60 Garden St., Cambridge, MA 02138, USA}
\email{weicheng.zang@cfa.harvard.edu}
\author{Hongjing Yang}
\affiliation{Department of Astronomy, Westlake University, Hangzhou 310030, Zhejiang Province, China}
\affiliation{Department of Astronomy, Tsinghua University, Beijing 100084, China}
\email{yang-hj19@mails.tsinghua.edu.cn}
\author{Doeon Kim}
\affiliation{Department of Physics, Chungbuk National University, Cheongju 28644, Republic of Korea}
\email{qso21@hanmail.net}
\author{Dong-Jin Kim}
\affiliation{Korea Astronomy and Space Science Institute, Daejon 34055, Republic of Korea}
\email{keaton03@kasi.re.kr}
\author{Byeong-Gon Park}
\affiliation{Korea Astronomy and Space Science Institute, Daejon 34055, Republic of Korea}
\email{bgpark@kasi.re.kr}
\collaboration{14}{(KMTNet Collaboration)}
\author{Przemek Mr{\'o}z}
\affiliation{Astronomical Observatory, University of Warsaw, Al.~Ujazdowskie 4, 00-478 Warszawa, Poland}
\email{pmroz@astrouw.edu.pl}
\author{Micha{\l} K. Szyma{\'n}ski}
\affiliation{Astronomical Observatory, University of Warsaw, Al.~Ujazdowskie 4, 00-478 Warszawa, Poland}
\email{msz@astrouw.edu.pl}
\author{Jan Skowron}
\affiliation{Astronomical Observatory, University of Warsaw, Al.~Ujazdowskie 4, 00-478 Warszawa, Poland}
\email{jskowron@astrouw.edu.pl}
\author{Rados{\l}aw Poleski} 
\affiliation{Astronomical Observatory, University of Warsaw, Al.~Ujazdowskie 4, 00-478 Warszawa, Poland}
\email{radek.poleski@gmail.co}
\author{Igor Soszy{\'n}ski}
\affiliation{Astronomical Observatory, University of Warsaw, Al.~Ujazdowskie 4, 00-478 Warszawa, Poland}
\email{soszynsk@astrouw.edu.pl}
\author{Pawe{\l} Pietrukowicz}
\affiliation{Astronomical Observatory, University of Warsaw, Al.~Ujazdowskie 4, 00-478 Warszawa, Poland}
\email{pietruk@astrouw.edu.pl}
\author{Szymon Koz{\l}owski} 
\affiliation{Astronomical Observatory, University of Warsaw, Al.~Ujazdowskie 4, 00-478 Warszawa, Poland}
\email{simkoz@astrouw.edu.pl}
\author{Krzysztof A. Rybicki}
\affiliation{Astronomical Observatory, University of Warsaw, Al.~Ujazdowskie 4, 00-478 Warszawa, Poland}
\affiliation{Department of Particle Physics and Astrophysics, Weizmann Institute of Science, Rehovot 76100, Israel}
\email{krybicki@astrouw.edu.pl}
\author{Patryk Iwanek}
\affiliation{Astronomical Observatory, University of Warsaw, Al.~Ujazdowskie 4, 00-478 Warszawa, Poland}
\email{piwanek@astrouw.edu.pl}
\author{Krzysztof Ulaczyk}
\affiliation{Department of Physics, University of Warwick, Gibbet Hill Road, Coventry, CV4 7AL, UK}
\email{kulaczyk@astrouw.edu.pl}
\author{Marcin Wrona}
\affiliation{Astronomical Observatory, University of Warsaw, Al.~Ujazdowskie 4, 00-478 Warszawa, Poland}
\affiliation{Villanova University, Department of Astrophysics and Planetary Sciences, 800 Lancaster Ave., Villanova, PA 19085, USA}
\email{mwrona@astrouw.edu.pl}
\author{Mariusz Gromadzki}          
\affiliation{Astronomical Observatory, University of Warsaw, Al.~Ujazdowskie 4, 00-478 Warszawa, Poland}
\email{marg@astrouw.edu.pl}
\author{Mateusz J. Mr{\'o}z} 
\affiliation{Astronomical Observatory, University of Warsaw, Al.~Ujazdowskie 4, 00-478 Warszawa, Poland}
\email{mmroz@astrouw.edu.pl}
\collaboration{100}{(The OGLE Team)}

\begin{abstract}
We investigated microlensing events detected by the OGLE and KMTNet surveys during the
2024 observing season, focusing on those that exhibit very complex anomaly features. 
Through this analysis, we found that the light curves of three events including 
OGLE-2024-BLG-0657, KMT-2024-BLG-2017, and KMT-2024-BLG-2480 cannot be readily 
interpreted using standard three-body lensing models such as a binary lens with 
a single source (2L1S) or a single lens with a binary source (1L2S).  In this work 
we present detailed analyses of these events to uncover the nature of their anomalous 
features.
An initial analysis using 2L1S modeling of the light curves showed that while it was 
difficult to simultaneously explain all of the multiple anomaly features, the main 
anomaly feature could be accounted for. Based on this model, we conducted four-body 
modeling that includes an additional lens or source. Through this approach, we found 
that the complex anomalies observed in the three events could be explained by a 2L2S 
model, in which both the lens and the source are binaries.
Analysis of the color and magnitude revealed that the source is a binary system consisting 
of G- and K-type main sequence stars for OGLE-2024-BLG-0657, two K-type main sequence stars
for KMT-2024-BLG-2017, and a K-type star with an early G-type main sequence companion for
KMT-2024-BLG-2480. A Bayesian analysis incorporating constraints from the lensing
observables indicates that the lenses in KMT-2024-BLG-2017 and KMT-2024-BLG-2480 are
likely binary systems of low-mass stars located in the Galactic bulge, whereas the lens 
system OGLE-2024-BLG-0657L is likely a binary composed of two stellar remnants situated 
in the Galactic disk.
\end{abstract}

\keywords{Gravitational microlensing (672)}

\section{Introduction} \label{sec:one}

The Korea Microlensing Telescope Network \citep[KMTNet;][]{Kim2016} is an 
observational survey that has been monitoring stars toward the Galactic bulge 
since 2015 to detect gravitational microlensing events. Its primary scientific 
goal is to identify short-duration planetary signals in microlensing light 
curves, aiming to uncover a large population of exoplanets located beyond 
the snow line.  In conjunction with other major surveys, such as the Optical 
Gravitational Lensing Experiment \citep[OGLE;][]{Udalski2015} and the Microlensing 
Observations in Astrophysics program \citep[MOA;][]{Bond2001, Sumi2003}, the current 
microlensing surveys collectively detects more than 4,000 events each year.

The vast majority of lensing events are modeled by the light curve of a single-lens 
single-source (1L1S) configuration. In such events, the magnification as a function 
of time is described by
\begin{equation}
A(t) = \frac{u^2 + 2}{u \sqrt{u^2 + 4}}; \qquad
u(t) = \sqrt{u_0^2 + \left( \frac{t - t_0}{t_{\rm E}} \right)^2},
\label{eq1}
\end{equation}
where $u(t)$ is the projected separation between the lens and source in units of the 
angular Einstein radius $\theta_{\rm E}$. The parameter $u_0$ represents the minimum 
separation (impact parameter), which occurs at time $t_0$, the moment of closest 
alignment. The quantity $t_{\rm E}$ is the Einstein timescale, defining the 
characteristic duration of the event.  This form yields a light curve that is smooth 
and symmetric about $ t_0$ \citep{Paczynski1986}.

For a small fraction of microlensing events, the observed light curves deviate from the 
standard form expected for a 1L1S event. In most of these anomalous cases, the deviations 
can be well explained by models in which three bodies (lens and source combined) are 
involved in the lensing system. These include scenarios in which the lens is composed 
of two masses, referred to as a binary-lens single-source (2L1S) event, or cases for 
which the source is a binary system, referred to as a single-lens binary-source (1L2S) 
event.

In a 2L1S event, the presence of two lens masses leads to the formation of caustics 
on the source plane. These are closed curves along which the magnification of a point 
source diverges to infinity.  When the source star passes over or near these caustic 
structures, the magnification changes rapidly, resulting in characteristic anomalies 
in the light curve, such as sharp spikes, bumps, or dips, depending on the source 
trajectory relative to the caustics \citep{Schneider1986, Mao1991}.

By contrast, a 1L2S event involves a single lens magnifying two distinct source stars. 
Each source is magnified independently according to its own position with respect to 
the lens, and the total observed flux is the sum of the two magnified fluxes. As a 
result, the light curve deviates from the smooth, symmetric shape characteristic of 
a 1L1S event, often displaying asymmetries or multiple peaks that reflect the properties 
and alignment of the two sources \citep{Griest1992}.

In rare cases, microlensing light curves exhibit highly complex deviations that cannot 
be fully explained by a three-body lensing system. For many of these events, accounting 
for the observed anomalies requires a four-body lensing model, involving either an 
additional lens component to a 2L1S system, leading to a triple-lens single-source (3L1S) 
configuration, or an additional source component, resulting in a binary-lens binary-source 
(2L2S) event.  An exceptional case is OGLE-2015-BLG-1459, in which the event was best 
explained by a single lens acting on three distinct source stars \citep{Hwang2018}.

Triple-lens systems produce highly intricate caustic structures that can consist of 
multiple disconnected, nested, or self intersecting curves \citep{Danek2015a, Danek2015b, 
Danek2019}.  These caustic topologies can be broadly categorized into two types. In 
``resonant'' triple-lens systems, in which the three lens components have comparable 
masses and are separated by distances on the order of the Einstein radius, the resulting 
caustics are extremely complex, often exhibiting overlapping or nested patterns. In 
contrast, ``hierarchical'' triple-lens systems, in which the third body is either a 
low mass planet or a distant stellar companion, produce caustics that resemble those 
of a dominant binary-lens system with a small perturbation from the third body 
\citep{Bozza1999, Han2001, Han2005}.

\begin{deluxetable*}{lcc}
\tablecaption{Events and coordinates.\label{table:one}}
\tablecolumns{3}
\tablewidth{0pt} 
\tablehead{
\colhead{Event} & \colhead{(${\rm RA},{\rm DEC}$)$_{2000}$} & \colhead{($l, b$)}
}
\startdata
OGLE-2024-BLG-0657 & (17:51:52.05, -23:21:26.2)   & $(5^\circ\hskip-2pt .5040$,   1$^\circ\hskip-2pt .6738)$ \\
KMT-2024-BLG-2017  & (18:03:33.96, -29:55:49.80)  & $(1^\circ\hskip-2pt .1149$,  -3$^\circ\hskip-2pt .8746)$ \\
KMT-2024-BLG-2480  & (17:57:06.01, -30:45:33.52)  & (-$0^\circ\hskip-2pt .2973$, -3$^\circ\hskip-2pt .0710)$ \\
\enddata
\end{deluxetable*}

A total of 15 microlensing events involving 3L1S configurations have been identified to 
date.  Among these, seven events were caused by planetary systems composed of two planets 
orbiting a single host star. These events are: OGLE-2006-BLG-109 \citep{Gaudi2008}, 
OGLE-2012-BLG-0026 \citep{Han2013}, OGLE-2018-BLG-1011 \citep{Han2019}, OGLE-2019-BLG-0468 
\citep{Han2022a}, KMT-2021-BLG-1077 \citep{Han2022b}, KMT-2021-BLG-0240 \citep{Han2022d}, 
and KMT-2022-BLG-1818 \citep{Li2025}. Another seven events were produced by systems in 
which a planet orbits one component of a binary stellar system. These include: 
OGLE-2006-BLG-284 \citep{Bennett2020}, OGLE-2007-BLG-349 \citep{Bennett2016}, 
OGLE-2008-BLG-092 \citep{Poleski2014}, OGLE-2016-BLG-0613 \citep{Han2017}, 
OGLE-2018-BLG-1700 \citep{Han2020}, KMT-2020-BLG-0414 \citep{Zang2021}, and 
OGLE-2023-BLG-0836 \citep{Han2024a}. The remaining event, KMT-2021-BLG-1122 
\citep{Han2023b}, was generated by a triple stellar system.

Lensing systems involving two lens masses and two source stars also produce very complex 
light curves. In such cases, each source is magnified independently according to its own 
position relative to the binary lens, and the observed light curve is the sum of the 
fluxes from the two magnified source stars.  Even with a single source, caustic-induced 
anomalies can create complex light curve structures, which become even more intricate 
when additional distortions from a secondary source are superimposed.

To date, 14 lensing events characterized by 2L2S configurations have been reported. These 
events include: MOA-2010-BLG-117 \citep{Bennett2018}, OGLE-2016-BLG-0882, OGLE-2017-BLG-0448 
\citep{Shin2024}, OGLE-2016-BLG-1003 \citep{Jung2017}, KMT-2018-BLG-1743 \citep{Han2021b}, 
OGLE-2018-BLG-0584, KMT-2018-BLG-2119 \citep{Han2023a}, KMT-2019-BLG-0797 \citep{Han2021a}, 
KMT-2021-BLG-1547 \citep{Han2023c}, KMT-2021-BLG-1898 \citep{Han2022c}, KMT-2021-BLG-0284, 
KMT-2022-BLG-2480, KMT-2024-BLG-0412 \citep{Han2024b}, and KMT-2022-BLG-0086 \citep{Chung2025}.

This paper presents the analysis of three additional 2L2S microlensing events discovered 
during the 2024 observing season: OGLE-2024-BLG-0657, KMT-2024-BLG-2017, and KMT-2024-BLG-2480.  
These events show complex light curves with multiple anomalies that cannot be explained by 
a conventional three-body model and instead require a four-body interpretation with two 
lens masses and two source stars.  For each event, we investigate the light curve features 
that reveal the binarity of both the lens and the source. We then provide a detailed 
description of the modeling process used to describe the observed light curves. Based 
on the modeling results, we determine the characteristics of the lens and source systems.

\section{Observations and data} \label{sec:two}

The 2L2S nature of the events analyzed in this work was identified through an investigation 
of microlensing events discovered by the OGLE and KMTNet surveys during the 2024 observing
season. Over this period, the two surveys detected a total of 1,389 and 3,441 events, 
respectively, with a substantial number of overlapping detections.

We began our analysis by modeling the events under the 1L1S assumption and selected those 
that showed deviations from this model. Approximately one-tenth of the events exhibited 
such anomalies. For these anomalous cases, we applied three-body modeling to identify the 
origin of the deviations. In the vast majority of cases, the anomalies were well explained 
by either a 2L1S or 1L2S model.

However, for a small subset of events, the observed light curves could not be adequately 
described by three-body models. For these cases, we carried out four-body modeling, 
taking into account the specific features that could not be explained by the three-body 
models. This analysis revealed that four events required a four-body model, specifically 
a configuration in which both the lens and the source are binary systems, to account for 
the observed anomalies. Among these, the analysis of KMT-2024-BLG-0412 was presented by 
\citep{Han2024b}, while the results for the remaining three events are reported in this 
work. Table~\ref{table:one} lists the event IDs along with their equatorial and Galactic 
coordinates. Among the three events, OGLE-2024-BLG-0657 was identified by the OGLE survey, 
while KMT-2024-BLG-2017 and KMT-2024-BLG-2480 were discovered by the KMTNet survey.

The photometric data for the events were acquired through observations conducted with 
instruments operated by the two survey groups.  The OGLE survey operates a telescope 
located at Las Campanas Observatory in Chile. The telescope has a 1.3-meter aperture 
and is equipped with a camera that covers a 1.4 square-degree field of view. The KMTNet 
survey employs three identical telescopes that are strategically distributed across three 
sites in the Southern Hemisphere: the Cerro Tololo Inter-American Observatory in Chile 
(KMTC), the South African Astronomical Observatory in South Africa (KMTS), and the Siding 
Spring Observatory in Australia (KMTA). Each KMTNet telescope has a 1.6-meter aperture 
and the field of view of the camera is 4 square degrees.

For both surveys, primary observations were conducted in the $I$ band, with approximately 
10\% the images taken in the $V$ band to enable color measurements.  Image reduction and
photometry of the events were performed using pipelines developed by the respective survey
groups. The KMTNet pipeline was developed by \citet{Albrow2009}, while the OGLE pipeline
was developed by \citet{Udalski2003}. Both pipelines are based on algorithms that implement 
the difference imaging method \citep{Tomaney1996, Alard1998}. To ensure consistency across 
data sets obtained from different telescopes, we normalized the error bars such that the
errors reflect the actual scatter in the data, and the reduced chi-square value of the 
best-fit model for each data set is equal to unity. This errorbar normalization process 
was done following the routine detailed in \citet{Yee2012}.

\section{Model lensing light curve} \label{sec:three}

Lensing light curves with complex anomaly features are often associated with multiple 
components in the lens or source system. Modeling such light curves requires a large 
set of parameters to describe their structure accurately. In this section, we present 
the lensing parameters used to characterize light curves arising from various lens-system 
configurations.

\subsection{Three-body lensing events} \label{sec:three-one}

Among lensing events involving a total of three bodies, the 1L2S and 2L1S configurations
correspond to cases in which an additional source or lens component, respectively, is added 
to the standard 1L1S scenario. Accordingly, additional parameters are required to describe 
the extra source or lens component compared to a 1L1S event.

For 1L2S events, in addition to the parameters describing the primary source component $S_1$, 
it is necessary to introduce additional parameters to characterize the second source component 
$S_2$: $t_{0,2}$, $u_{0,2}$, and $q_F$. Here, $t_{0,2}$ represents the time of closest approach 
of $S_2$ to the lens, and $u_{0,2}$ is its corresponding impact parameter. The parameter $q_F$ 
denotes the flux ratio between the two sources, defined as $q_F = F_{S,2} / F_{S,1}$, where 
$F_{S,1}$ and $F_{S,2}$ are the fluxes of $S_1$ and $S_2$, respectively.  We designate the 
source that approaches closer to the lens as $S_1$, and therefore $q_F$ can be greater than 
unity. With these parameters, the total magnification of a 1L2S event is given by
\begin{equation}
A(t) = { A_1(t) + q_FA_2(t) \over 1+q_F},
\label{eq2}
\end{equation}
where $A_1(t)$ and $A_2(t)$ are the magnifications of the individual source components $S_1$
and $S_2$, respectively. To distinguish the 1L1S parameters of the first source from those of 
the second, we denote them as $(t_{0,1}, u_{0,1})$.

For 2L1S events, the additional parameters needed to describe the second lens component are 
$s$, $q$, and $\alpha$. The parameters $s$ and $q$ represent the projected separation and mass
ratio between the two lens components, $M_1$ and $M_2$, respectively. We designate $M_1$ as
the component responsible for the major anomaly in the light curve, and thus $q$ can be greater
than unity. The separation $s$ is normalized to the angular Einstein radius. Although not directly
related to the lens-system configuration, an additional parameter required in the 2L1S model is 
the normalized source radius $\rho$, defined as the ratio of the angular source radius $\theta_*$ 
to $\thetae$.  This parameter is required to model the light curve during caustic crossings, 
where finite-source effects influence the magnification.

\subsection{Four-body lensing events} \label{sec:three-two}

Among events involving four bodies, 2L2S and 3L1S events correspond to cases in which one
additional source and one additional lens component, respectively, are added to a 2L1S event.
Accordingly, modeling these events requires additional parameters to describe the extra source 
or lens component beyond those used in the 2L1S configuration.

For 3L1S events, additional parameters are introduced to describe the third lens component 
($M_3$). These include $s_3$, $q_3$, and $\psi$, where $s_3$ is the projected separation 
between $M_3$ and the primary lens component $M_1$, $q_3$ is their mass ratio, and $\psi$ 
denotes the orientation angle of $M_3$, measured with respect to the $M_1$--$M_2$ axis. 
The subscript ``3'' designates parameters associated with the third lens, while those 
describing the second lens component $M_2$ are denoted with the subscript ``2''.

In the case of 2L2S events, the parameters used to describe the additional source are the 
same as those for 1L2S events, namely $t_{0,2}$, $u_{0,2}$, and $q_F$. Because each source 
in a 2L2S event is likely to cross a caustic, it is necessary to account for the finite-source 
effects that occur during these crossings. Therefore, the normalized source radii corresponding 
to each source, $\rho_1$ and $\rho_2$, are additionally required.

If the lens is a binary system, the orbital motion of the lens can induce distortions in
the lensing light curve. \citet{Nucita2014} pointed out that an accurate timing analysis
of the residuals from the model, such as using a Lomb-Scargle periodogram, can help
infer the orbital period of the binary lens. To investigate the potential impact of orbital
effects in the analyzed events, we examined the residuals from each event's best-fit
model. However, we did not identify any noticeable deviations attributable to lens orbital
motion during the course of the events. Therefore, we did not consider lens orbital
motion as an additional component in the modeling.

\section{Analyses of events} \label{sec:four}

In this section, we describe the modeling procedures for the individual events and present 
the corresponding lensing solutions. We also show the configurations of the lens systems 
and explain how the anomaly features in the light curves are produced.

\begin{figure}[t]
\includegraphics[width=\columnwidth]{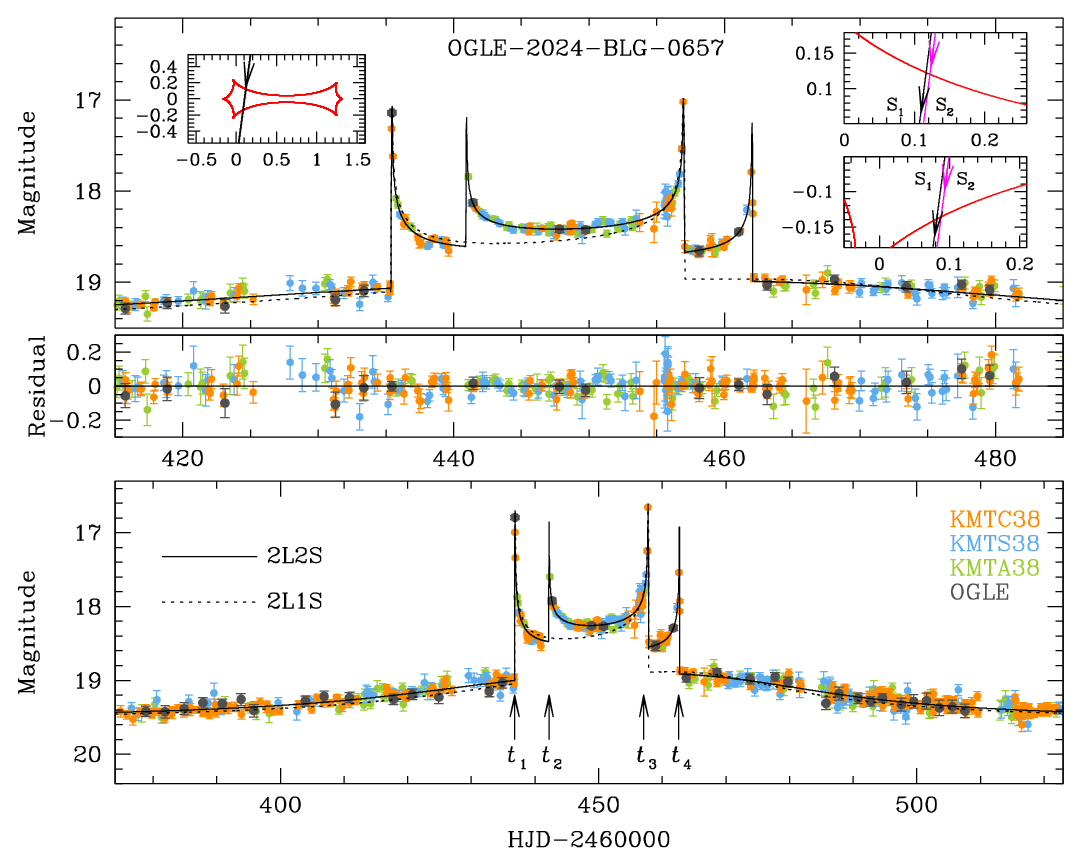}
\caption{
Light of lensing event OGLE-2024-BLG-0657.  The lower panel displays the full light curve, 
while the upper panel provides a zoomed-in view around the anomaly. The left inset of the 
upper panel illustrates the lens system configuration, and the two right insets show 
zoomed-in views at the times of the source's caustic entrance and exit. The black and 
magenta arrowed lines indicate the trajectories of the source stars $S_1$ and $S_2$, 
respectively. The four arrows labeled $t_1$, $t_2$, $t_3$, and $t_4$ in the lower panel 
indicate the times of the four caustic crossings. The colors of the labels in the legend 
correspond to those of the data points.
}
\label{fig:one}
\end{figure}

\subsection{OGLE-2024-BLG-0657} \label{sec:four-one}

The microlensing event OGLE-2024-BLG-0657 was discovered by the OGLE collaboration on 2024 
June 2, corresponding to the abridged heliocentric Julian date ${\rm HJD}^\prime \equiv 
{\rm HJD} - 2460000 = 463$. The source star exhibited a baseline magnitude of $I_{\rm base} 
= 19.51$ in the $I$-band. Significant extinction was present along the line of sight, with 
an $I$-band extinction of $A_I = 3.12$. Although the event was not independently identified 
by the KMTNet survey during the 2024 observing season, it was subsequently recovered through 
postseason photometric analysis of the OGLE-discovered source. The source was located in 
KMTNet field BLG04, which was observed with a cadence of 1.0 hour.

\begin{figure}[t]
\includegraphics[width=\columnwidth]{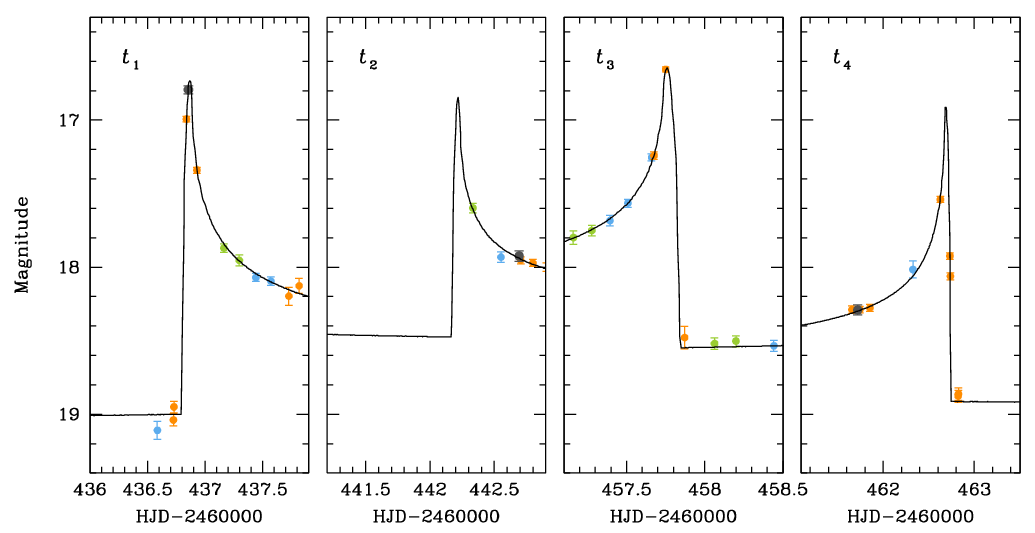}
\caption{
Enlarged views of the individual caustic spikes in the light curve of OGLE-2024-BLG-0657.
}
\label{fig:two}
\end{figure}

Figure~\ref{fig:one} displays the lensing light curve of OGLE-2024-BLG-0657.  It reveals 
a highly complex pattern featuring four spikes at ${\rm HJD}^\prime \sim 436.8$ ($t_1$), 
442.2 ($t_2$), 457.1 ($t_3$), and 462.6 ($t_4$). The spikes at $t_1$ and $t_2$ show a 
sharp increase in magnification followed by a more gradual decline, a characteristic 
signature of a caustic entry. In contrast, the spikes at $t_3$ and $t_4$ display the 
opposite trend, with a slow rise in magnification followed by an abrupt drop, consistent 
with a caustic exit.  Figure~\ref{fig:two} presents enlarged views of the individual 
caustic spikes.

The observed light curve cannot be explained by a 2L1S model. A binary-lens system can 
generate one or more sets of caustics, depending on the projected separation between the two 
lens components.  Each caustic forms a closed curve, and as the source star moves through the 
caustic, a pair of spikes are observed, with one occurring when the source enters the caustic 
and another when it exits.  However, the light curve shown in Figure~\ref{fig:one} displays 
two consecutive caustic-entry spikes followed by two consecutive caustic-exit spikes. This 
pattern is inconsistent with the expected behavior of a 2L1S event, in which every caustic-entry 
spike must be followed by a corresponding exit spike. This discrepancy indicates that the anomaly 
cannot be explained within the 2L1S framework and requires a more complex model.

An anomaly exhibiting two overlapping pairs of caustic-crossing spikes can arise through two
distinct pathways. The first occurs when the lens system consists of three masses. In such 
cases, the caustics from the three lens components can interact, producing intersecting or 
twisted structures. As the source star passes through these intersecting caustic regions, the 
light curve may display overlapping pairs of caustic-crossing features. An example of this 
type of anomaly, characterized by entangled caustic-crossing signals, is found in the light 
curve of the 3L1S event OGLE-2016-BLG-0613 \citep{Han2017}.

Another pathway for producing overlapping caustic-crossing features arises when both the 
lens and source are binaries. In this case, one pair of caustic-crossing features can occur 
as one source component crosses a caustic, while a separate pair is produced by the other 
source crossing a different part of the caustic structure. If the two pairs of caustic 
crossings occur close in time, their features can overlap in the light curve, resulting 
in a overlapping caustic-crossing anomaly.  The light curve of the event KMT-2021-BLG-0284 
\citep{Han2024b} illustrates such a case.

The four-body modeling process was initiated using the 2L2S configuration. Because a 2L2S 
light curve represents the combined magnification patterns of two source stars, the modeling 
was performed in two stages. In the first stage, we identified a 2L1S model that reproduced 
one pair of caustic-crossing features. In the second stage, a second source was introduced 
to account for the remaining pair of caustic crossings not explained by the initial model. 
This two-step approach demonstrated that the observed light curve can be effectively 
interpreted as the superposition of two distinct 2L1S events: one generating the caustic 
spikes at $t_1$ and $t_3$, and the other accounting for those at $t_2$ and $t_4$.  As 
illustrated in Figure~\ref{fig:one}, the dotted curve represents the 2L1S model reproducing 
the features at $t_1$ and $t_3$, while the solid curve shows the full 2L2S model that captures 
all caustic structures, including those at $t_2$ and $t_4$.

\begin{deluxetable}{ll}
\tablecaption{Best-fit parameters of OGLE-2024-BLG-0657.\label{table:two}}
\tablecolumns{2}
\tablewidth{0pt} 
\tablehead{
\colhead{Parameter} & \colhead{Value}
}
\startdata
$\chi^2$                   & $973.8$              \\
$t_{0,1}$ (HJD$^\prime$)   & $447.719 \pm 0.071$  \\
$u_{0,1}$                  & $0.0994 \pm 0.0046$  \\
$t_{0,2}$ (HJD$^\prime$)   & $452.950 \pm 0.078$  \\
$u_{0,2}$                  & $0.1052 \pm 0.0049$  \\
$\te$ (days)               & $79.41 \pm 1.91$     \\
$s$                        & $1.766 \pm 0.020$    \\
$q$                        & $1.610 \pm 0.067$    \\
$\alpha$ (rad)             & $4.8449 \pm 0.0080$  \\
$\rho_1$ ($10^{-3}$)       & $0.555 \pm 0.032$    \\
$\rho_2$ ($10^{-3}$)       & $0.397 \pm 0.043$    \\
$q_F$                      & $0.677 \pm 0.015$    \\
\enddata
\tablecomments{${\rm HJD}^\prime = {\rm HJD}-2460000$.}
\end{deluxetable}

We also explored a 3L1S interpretation, assuming the lens system to be a hierarchical triple. 
In this approach, the parameters of the 2L1S model that explain the caustic features at $t_1$ 
and $t_3$ were kept fixed, while a grid search was carried out over the parameters describing 
the third lens component  ($s_3$, $q_3$, and $\psi$) in order to account for the remaining 
features. However, the 3L1S modeling did not yield a solution that provides a fit comparable 
to that of the 2L2S model.

Table~\ref{table:two} presents the best-fit parameters of the 2L2S solution. The binary-lens 
parameters are $(s, q) \sim (1.77, 1.61)$, indicating that the lens consists of two components 
with roughly equal masses and a projected separation greater than the Einstein radius. This 
configuration produces a resonant caustic elongated along the binary axis.  The source is 
composed of two stars with a flux ratio of approximately $q_F \sim 0.68$ between them.  The 
normalized source radius of $S_1$ was determined from the partially resolved caustic crossings 
at $t_1$, while that of $S_2$ was derived from the caustic feature at $t_4$.

The lens system configuration is illustrated in the insets of the upper panel of Figure 
\ref{fig:one}.  Both source stars crossed the left side of the caustic at an incidence angle 
of approximately $82^\circ$.  The small difference in their impact parameters, $\Delta u_0 
= u_{0,2} - u_{0,1} = 0.0068$, indicates that the two sources followed nearly identical 
trajectories. The difference in the times of closest approach, $\Delta t_0 = t_{0,2} - 
t_{0,1} = 5.23$ days, shows that $S_2$ trailed $S_1$ by about 5.2 days. The caustic features 
at $t_1$ and $t_3$ were produced when $S_1$ entered and exited the caustic, respectively, 
while the features at $t_2$ and $t_4$ correspond to the entry and exit of $S_2$. The time 
interval between the caustic spikes at $t_1$ and $t_3$, $\Delta t_{\rm c} \sim 20.3$ days, 
is significantly longer than the source time offset $\Delta t_0 \sim 5.23$ days. Consequently, 
the spike from the caustic entry of $S_2$ occurred before the caustic exit of $S_1$, leading 
to an overlap of the two sets of caustic-induced spikes from $S_1$ and $S_2$.

\begin{figure}[t]
\includegraphics[width=\columnwidth]{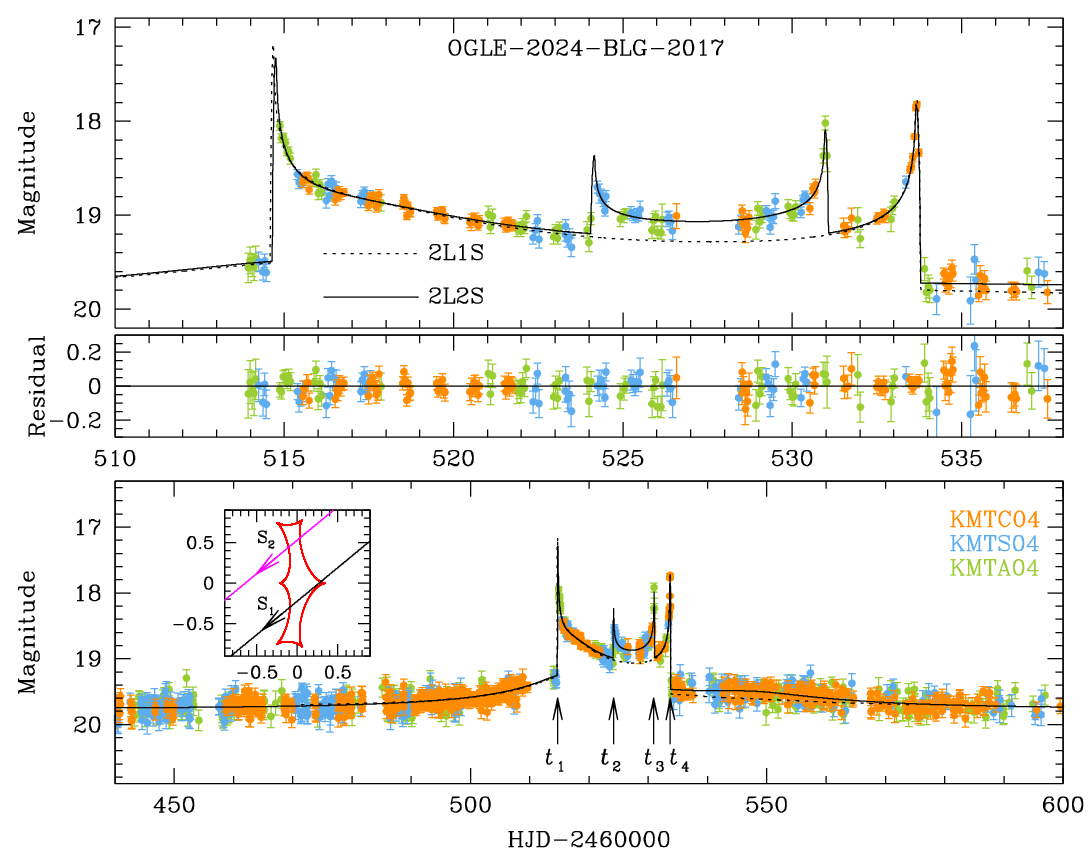}
\caption{
Light of lensing event KMT-2024-BLG-2017.  Notations are same as those in Fig.~\ref{fig:one}.
}
\label{fig:three}
\end{figure}

\subsection{KMT-2024-BLG-2017} \label{sec:four-two}

The microlensing event KMT-2024-BLG-2017 was detected by the KMTNet survey on 2024 July 
31, corresponding to ${\rm HJD}^\prime = 522$. The source star, which has a baseline 
magnitude of $I_{\rm base} = 19.98$, is situated in the KMTNet BLG04 field, along a line 
of sight with an $I$-band extinction of $A_I = 0.78$. Observations of this field were 
conducted at a cadence of 1.0 hour.

\begin{figure}[t]
\includegraphics[width=\columnwidth]{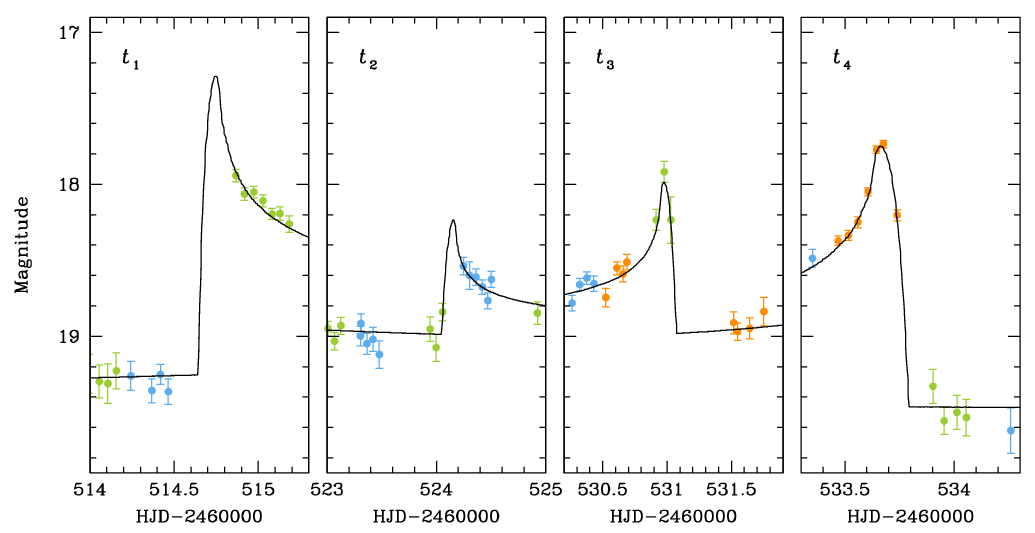}
\caption{
Enlarged views of the individual caustic spikes in the light curve of KMT-2024-BLG-2017.
}
\label{fig:four}
\end{figure}

The lensing light curve of the event is presented in Figure~\ref{fig:three}. Similar to 
the previous event OGLE-2024-BLG-0657, the light curve displays four distinct caustic 
spikes occurring at ${\rm HJD}^\prime \sim 514.7$ ($t_1$), $524.12$ ($t_2$), $530.93$ 
($t_3$), and $533.7$ ($t_4$).  Figure~\ref{fig:four} presents the zoomed-in views of the 
individual caustic spikes.  The U-shaped profiles between the caustic spikes indicate 
that the spikes at $t_2$ and $t_3$ form one caustic-crossing pair, while those at $t_1$ 
and $t_4$ constitute a second pair. In OGLE-2024-BLG-0657, the two caustic pairs appeared 
in an overlapping manner, whereas in KMT-2024-BLG-2017, a caustic pair with a short time 
interval is embedded within another pair separated by a longer time span.

We begin our analysis by modeling the light curve using a 2L2S configuration. As an 
initial step, we searched for a 2L1S solution that fits the light curve while excluding 
the data between $t_2$ and $t_3$. Subsequently, a second source star was introduced to 
explore a 2L2S solution. In Figure~\ref{fig:three}, the initial 2L1S model is represented 
by the dotted curve, while the final 2L2S model is shown as a solid curve.  This analysis 
reveals that the complex anomaly features in the light curve are well reproduced by the 
2L2S model.

The parameters of the 2L2S model are summarized in Table~\ref{table:three}. The binary-lens 
parameters are approximately $(s, q) \sim (0.89, 0.56)$, indicating that the event was 
produced by a binary lens composed of two nearly equal-mass components with a projected 
separation close to the Einstein radius. The event timescale is measured to be $\te \sim 
38$ days, and the flux ratio between the two source stars is $q_F \sim 0.74$. The partially 
resolved caustic crossing at $t_4$ enabled the measurement of $\rho_1$, while the resolved 
caustic features at $t_2$ and $t_3$ allowed the determination of $\rho_2$.

\begin{figure}[t]
\includegraphics[width=\columnwidth]{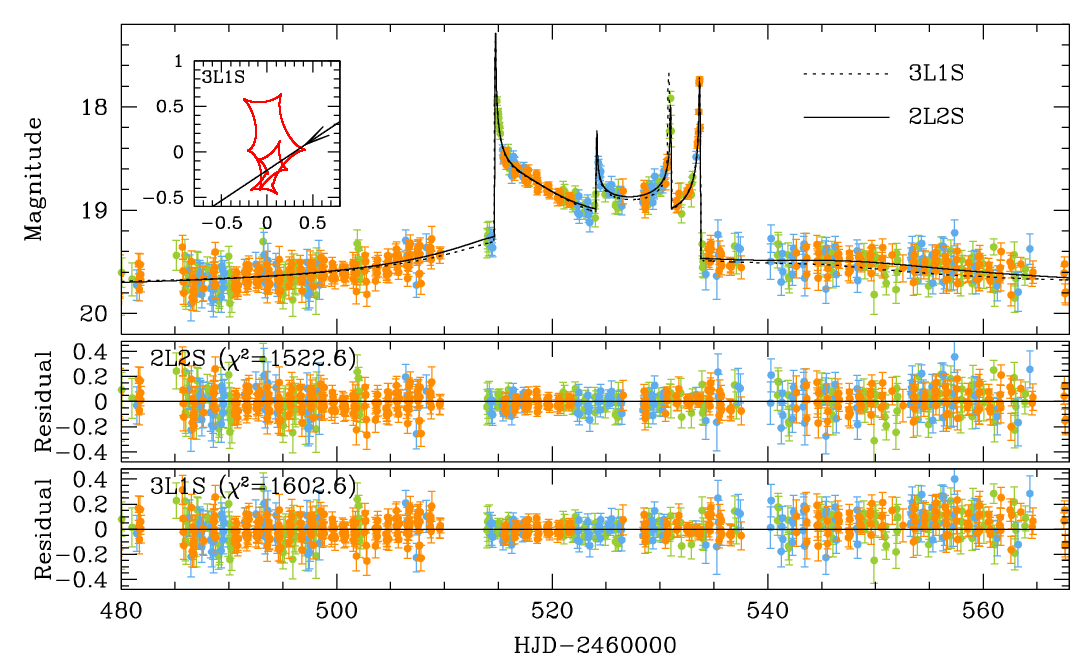}
\caption{
Comparison between the 2L2S and 3L1S model fits of KMT-2024-BLG-2017. The inset in the 
top panel is the lens-system configuration of the 3L1S model.
}
\label{fig:five}
\end{figure}

\begin{deluxetable}{lll}
\tablecaption{Best-fit parameters of KMT-2024-BLG-2017.\label{table:three}}
\tablecolumns{3}
\tablewidth{0pt} 
\tablehead{
\colhead{Parameter} & \colhead{2L2S} & \colhead{3L1S}
}
\startdata
$\chi^2$                   & $1522.6$             & $1602.8$             \\
$t_{0,1}$ (HJD$^\prime$)   & $523.70 \pm 0.21$    & $526.76 \pm 0.27$    \\
$u_{0,1}$                  & $0.1705 \pm 0.0039$  & $0.1678 \pm 0.0019$  \\
$t_{0,2}$ (HJD$^\prime$)   & $538.54 \pm 0.47$    & \nodata              \\
$u_{0,2}$                  & -$0.410 \pm 0.020$   & \nodata              \\
$\te$ (days)               & $38.46 \pm 0.77$     & $36.75 \pm 0.11$     \\
$s_2$                      & $0.8852 \pm 0.0046$  & $1.0000 \pm 0.0022$  \\
$q_2$                      & $0.575 \pm 0.029$    & $0.458 \pm 0.027$    \\
$\alpha$ (rad)             & $5.592 \pm 0.015$    & $5.695 \pm 0.017$    \\
$s_3$                      & \nodata              & $0.9749 \pm 0.0051$  \\
$q_3$                      & \nodata              & $0.0611 \pm 0.0018$  \\
$\psi$                     & \nodata              & $0.813 \pm 0.016$    \\
$\rho_1$ ($10^{-3}$)       & $1.64 \pm 0.08$      & $2.118 \pm 0.096$    \\
$\rho_2$ ($10^{-3}$)       & $1.44 \pm 0.37$      & \nodata              \\
$q_F$                      & $0.744 \pm 0.050$    & \nodata              \\
\enddata
\tablecomments{${\rm HJD}^\prime = {\rm HJD}-2450000$.}
\end{deluxetable}

The lens system configuration for the 2L2S model is displayed in the inset of the bottom 
panel of Figure~\ref{fig:three}. The binary lens generates a six-fold resonant caustic 
structure, elongated perpendicular to the binary axis. The first source crossed the caustic 
diagonally, entering through the upper-right fold and exiting through the lower-left fold, 
resulting in the spikes observed at $t_1$ and $t_4$. The second source followed the first, 
entering through the upper-right fold and exiting through the upper-left fold, which produced 
the spikes at $t_2$ and $t_3$.  Because the source passed through a region where the caustic 
is narrow, the separation between the caustic spikes caused by the second source is smaller 
than that caused by the first source.

We further explored the possibility that the anomaly could be explained by a 3L1S model. 
In this approach, we began with a grid search to identify the lensing parameters associated 
with the third body, while keeping the remaining parameters fixed to those of the 2L1S 
solution. We then refined the model by allowing all parameters to vary freely. The 
lensing parameters corresponding to the best-fit 3L1S solution are summarized in 
Table~\ref{table:three}, and the resulting model light curve, residuals, and lens-system 
configuration are presented in Figure~\ref{fig:five}.  The configuration reveals that 
the presence of the third lens mass significantly distorts the caustic, producing a 
twisted and nested structure. The source trajectory intersected this twisted region, 
generating the inner pair of caustic features that could not be reproduced by a 2L1S 
model.  According to the solution, the third lens component, which has a mass ratio of 
$q_3 \sim 0.06$ relative to the first component, is positioned at an orientation angle 
of $\psi \sim 46^\circ$.  While the 3L1S model provides a reasonable fit to the observed 
light curve, it is notably disfavored in comparison to the 2L2S model, with a substantial 
$\Delta\chi^2$ difference of 80.2.

\subsection{KMT-2024-BLG-2480} \label{sec:four-three}

The KMTNet group discovered the lensing event KMT-2024-BLG-2480 on 2024 September 9 
(${\rm HJD}^\prime = 562$), approximately two days before it reached a moderate peak 
magnification of $A \simeq 6.0$. The source star, with a baseline magnitude of $I_{\rm base} 
= 19.51$, is located in a region with an extinction of $A_I = 1.41$. The event occurred in 
the overlapping area of the KMTNet prime fields BLG01 and BLG41, which were observed at a 
cadence of 0.5 hours individually and 0.25 hours combined. Following the initial peak at 
${\rm HJD}^\prime = 564.5$ ($t_1$), the light curve gradually declined, then brightened 
again to a secondary peak at ${\rm HJD}^\prime = 586.9$ ($t_2$) with a magnification of 
$A \sim 2.2$ before returning to baseline.

\begin{figure}[t]
\includegraphics[width=\columnwidth]{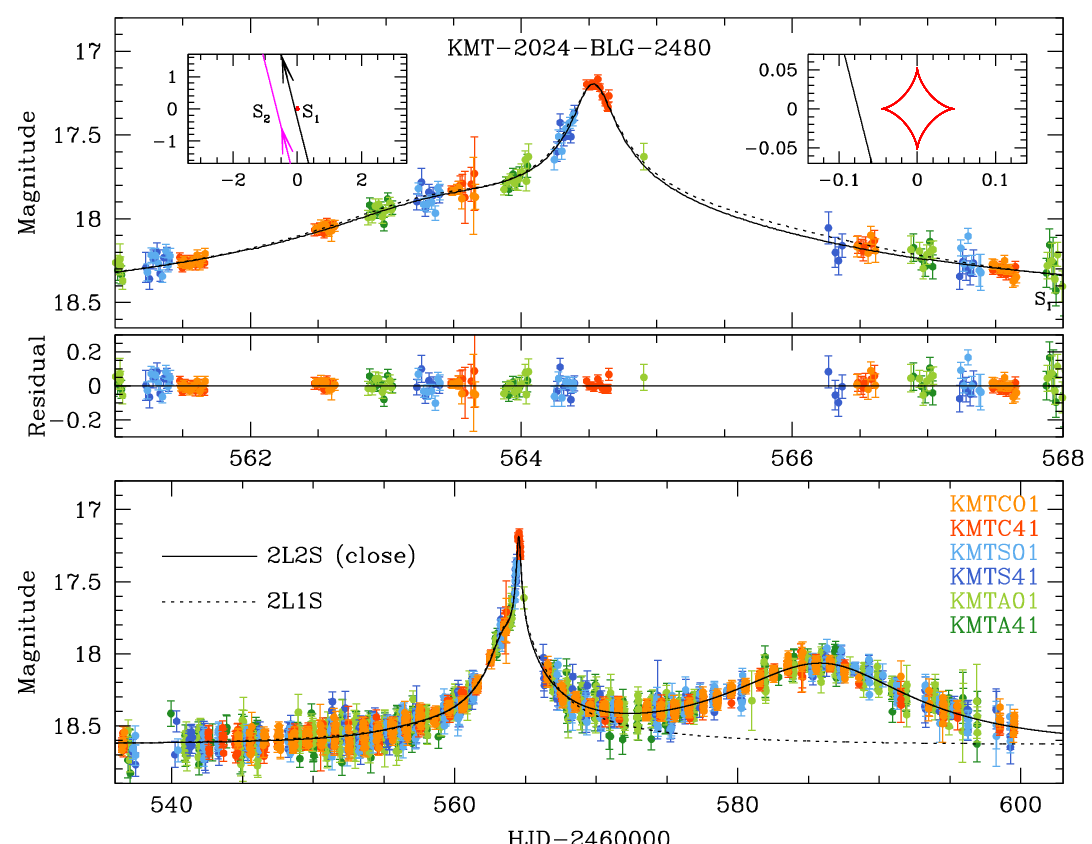}
\caption{
Light of lensing event KMT-2024-BLG-2480.
}
\label{fig:six}
\end{figure}

The lensing light curve for the event is presented in Figure~\ref{fig:six}. In addition 
to two distinct peaks, the light curve shows an extra deviation near the first peak. 
Modeling indicates that the observed features cannot be explained by any three-body 
configurations, including the 2L1S or 1L2S models. However, when the data after 
${\rm HJD}^\prime = 570$, which includes the second peak, are excluded, the remaining 
light curve is well described by a 2L1S model with a projected separation either much 
smaller (close model) or much larger (wide model) than the Einstein radius. This 2L1S 
model is plotted as a dotted curve over the data points in Figure~\ref{fig:six}.

Building on the 2L1S modeling that explains the first peak and the associated anomaly, 
we investigated a 2L2S model to account for the second bump. This analysis yielded two 
solutions corresponding to the close and wide binary-lens configurations. The lensing 
parameters for both 2L2S solutions, along with their corresponding $\chi^2$ values, are 
listed in Table~\ref{table:four}. The binary-lens parameters are $(s, q) \sim (0.31, 0.96)$ 
for the close solution and $(s, q) \sim (6.6, 9.4)$ for the wide solution. The degeneracy 
between the two solutions is strong, with the close solution being marginally favored by 
$\Delta\chi^2 = 3.0$. In addition to the large difference in separation $s$, the two 
solutions exhibit notably different time scales: $\te \sim 10$ days for the close solution 
and $\te \sim 44$ days for the wide solution. This discrepancy arises because the time scale 
of the wide solution is scaled to the total mass of the binary lens, $M_1 + M_2$, whereas 
the event duration is governed by $M_1$.  When the time scale of the wide solution is rescaled 
to $M_1$, it becomes $t_{{\rm E},1} = [1/(1+q)]^{1/2} \te \sim 13.6$ days, which is comparable 
to the time scale of the close solution.  For reference, the value of $t_{{\rm E},1}$ 
corresponding to the wide solution is also reported in Table~\ref{table:four}.

\begin{deluxetable}{lll}
\tablecaption{Best-fit parameters of KMT-2024-BLG-2480.\label{table:four}}
\tablecolumns{3}
\tablewidth{0pt} 
\tablehead{
\colhead{Parameter} & \colhead{Close} & \colhead{Wide}
}
\startdata
$\chi^2$                   & $2395.5$             & $2398.5$             \\
$t_{0,1}$ (HJD$^\prime$)   & $564.365 \pm 0.011$  & $564.302 \pm 0.013$  \\
$u_{0,1}$                  & $0.0734 \pm 0.0023$  & $0.0188 \pm 0.0015$  \\
$t_{0,2}$ (HJD$^\prime$)   & $585.904 \pm 0.063$  & $585.537 \pm 0.066$  \\
$u_{0,2}$                  & $0.623 \pm 0.036$    & $0.135 \pm 0.011$    \\
$\te$ (days)               & $10.03 \pm 0.38$     & $43.98 \pm 3.06$     \\
$t_{{\rm E},1}$ (days)     & \nodata              & $13.64 \pm 0.95$     \\
$s$                        & $0.3120 \pm 0.0042$  & $6.60 \pm 0.17$      \\
$q$                        & $0.955 \pm 0.045$    & $9.40 \pm 1.30$      \\
$\alpha$ (rad)             & $1.324 \pm 0.014$    & $1.284 \pm 0.015$    \\
$\rho_1$ ($10^{-3}$)       & \nodata              & \nodata              \\
$\rho_2$ ($10^{-3}$)       & \nodata              & \nodata              \\
$q_F$                      & $5.89 \pm 0.38$      & $4.06 \pm 0.16$      \\
\enddata
\tablecomments{${\rm HJD}^\prime = {\rm HJD}-2450000$.}
\end{deluxetable}

\begin{deluxetable*}{lllll}
\tablecaption{Source parameters, angular Einstein radii, and relative lens-source proper motions.\label{table:five}}
\tablehead{
\colhead{Parameter}          &
\colhead{OGLE-2024-BLG-0657} &
\colhead{KMT-2024-BLG-2017}  &
\colhead{KMT-2024-BLG-2480}
}
\startdata
$(V\!-\!I, I)_{S_1}$         & $(2.875 \pm 0.237, 20.557 \pm 0.010)$ & $(1.628 \pm 0.100, 21.629 \pm 0.016)$ & $(2.304 \pm 0.093, 20.731 \pm 0.034)$ \\
$(V\!-\!I, I)_{S_2}$         & $(3.814 \pm 0.632, 20.998 \pm 0.013)$ & $(1.904 \pm 0.166, 21.985 \pm 0.021)$ & $(1.978 \pm 0.063, 18.814 \pm 0.014)$ \\
$(V\!-\!I, I)_{\rm RGC}$     & $(3.174, 16.962)$                     & $(1.779, 16.156)$                     & $(2.373, 16.047)$ \\
$(V\!-\!I, I)_{{\rm RGC},0}$ & $(1.060, 14.204)$                     & $(1.060, 14.501)$                     & $(1.060, 14.501)$ \\
$(V\!-\!I, I)_{S_1,0}$       & $(0.761 \pm 0.237, 17.799 \pm 0.010)$ & $(0.909 \pm 0.100, 19.974 \pm 0.016)$ & $(0.991 \pm 0.093, 19.186 \pm 0.034)$ \\
$(V\!-\!I, I)_{S_2,0}$       & $(1.700 \pm 0.632, 18.240 \pm 0.013)$ & $(1.186 \pm 0.166, 20.330 \pm 0.021)$ & $(0.665 \pm 0.063, 17.269 \pm 0.014)$ \\
Type ($S_1$)                 & G6V                                   & K2V                                   & K3V \\
Type ($S_2$)                 & K7V                                   & K4V                                   & G0V \\
$\theta_{*,S_1}$ ($\mu$as)   & $0.919 \pm 0.227$                     & $0.398 \pm 0.049$                     & $0.630 \pm 0.074$ \\
$\theta_{*,S_2}$ ($\mu$as)   & $1.566 \pm 0.996$                     & $0.466 \pm 0.084$                     & $1.050 \pm 0.099$ \\
$\thetae$ (mas)              & $1.570 \pm 0.397$                     & $0.259 \pm 0.034$                     & \nodata \\
$\mu$ (mas/yr)               & $7.53 \pm 1.90$                       & $2.58 \pm 0.34$                       & \nodata \\
\enddata
\end{deluxetable*}

The lens system configuration for KMT-2024-BLG-2480 is shown in the insets of the upper 
panel of Figure~\ref{fig:six}. As discussed in Sect.~\ref{sec:five}, the wide solution 
is disfavored, and the illustrated configuration corresponds to the close solution. The 
binary lens produces a very small four-cusp caustic. The first source approached the left 
side of this caustic, resulting in the anomaly observed near the first peak of the light 
curve. The second peak was generated by a second source trailing the first with a time 
offset of $\Delta t = t_{0,2} - t_{0,1} \sim 21.5$ days. The flux ratio between the two 
sources is $q_F \sim 5.9$, indicating that $S_2$ is brighter than $S_1$. However, the 
impact parameter of $S_2$, $u_{0,2} \sim 0.62$, is much larger than that of $S_1$, 
$u_{0,1} \sim 0.07$, resulting in a lower magnification for the second peak despite the 
higher intrinsic brightness of $S_2$. Because neither $S_1$ nor $S_2$ crossed the caustic, 
their normalized source radii could not be constrained.

\section{Source stars and angular Einstein radii} \label{sec:five}

To determine the properties of the individual stars comprising the binary source in 
each event, we analyzed their reddening-corrected colors and magnitudes. As a first 
step, we measured the total flux from the binary source system, $F_{\rm S}$, by fitting 
the photometric data in the $V$- and $I$-bands to the best-fit microlensing model, 
$A_{\rm model}(t)$, using the relation:
\begin{equation}
F_{\rm obs}(t) = A_{\rm model}(t) F_{\rm S} + F_b,
\label{eq3}
\end{equation}
where $F_{\rm obs}(t)$ is the observed flux and $F_b$ denotes the contribution from 
unresolved, blended stars within the photometric aperture.  This flux measurement is 
based on photometry processed with the pyDIA reduction pipeline \citep{Albrow2017}.  
Given the total source flux and the flux ratio between the two source stars, the 
individual fluxes of the binary source components, $F_1$ and $F_2$, are computed as:
\begin{equation}
F_1 = \frac{1}{1 + q_F} F_{\rm S}, \qquad
F_2 = \frac{q_F}{1 + q_F} F_{\rm S}.
\label{eq4}
\end{equation}

\begin{figure}[t]
\includegraphics[width=\columnwidth]{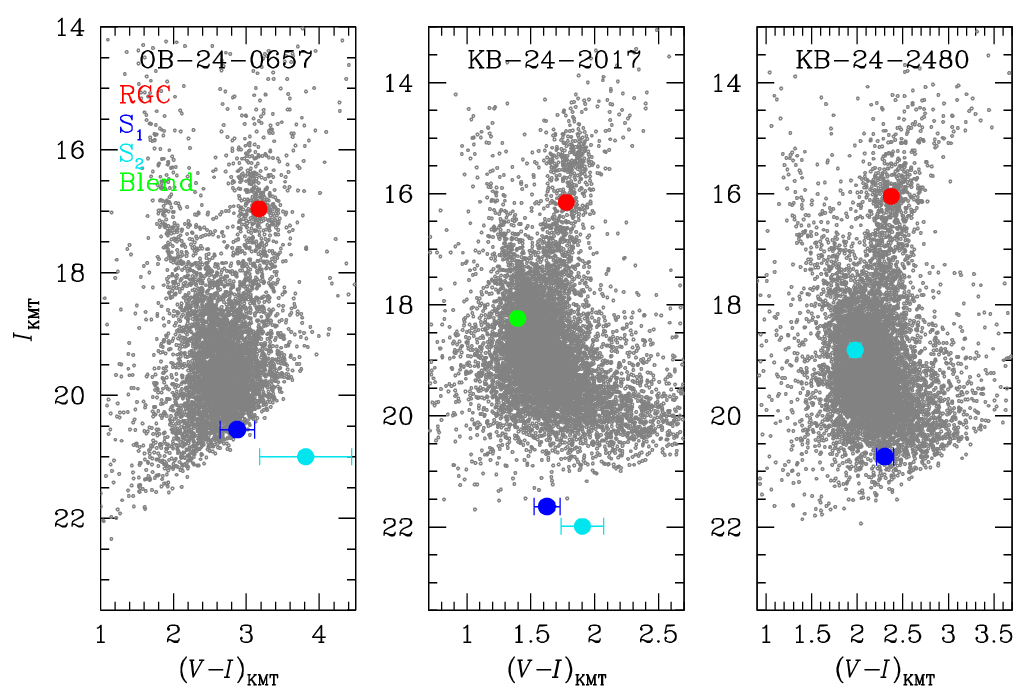}
\caption{
Locations of the binary source components ($S_1$ and $S_2$) in the instrumental 
color-magnitude diagram.  Also marked is the centroid of the red giant clump (RGC), 
which serves as the reference point for color and magnitude calibration.  For 
the event with measured blended flux, the blend position is marked.
}
\label{fig:seven}
\end{figure}

Figure~\ref{fig:seven} presents the positions of the source stars on instrumental 
color-magnitude diagrams (CMDs) of stars located in the vicinity of each source. The 
CMDs were generated using the same pyDIA photometry pipeline employed to measure the 
source fluxes. The instrumental (uncalibrated) colors and magnitudes of the two source 
components, $(V - I, I)_{{\rm S}_1}$ and $(V - I, I)_{{\rm S}_2}$, are provided in 
Table~\ref{table:five}.

In the second step, we calibrated the color and magnitude of the source stars. This 
calibration was performed by referencing the centroid of the red giant clump (RGC) 
in the CMD, following the methodology of \citet{Yoo2004}.  For the dereddened color 
and magnitude of the RGC, $(V - I, I)_{{\rm RGC},0}$, we adopted the values from 
\citet{Bensby2013} and \citet{Nataf2013}.  Table~\ref{table:five} presents the adopted 
RGC values, along with the dereddened colors and magnitudes of the two stars comprising 
the binary source, $(V - I, I)_{S_1,0}$ and $(V - I, I)_{S_2,0}$. Also included in the 
table are the spectral classifications of the binary source stars, inferred from their 
calibrated colors and magnitudes. Based on the color and magnitude, the source was 
identified as a binary composed of G- and K-type main-sequence stars for OGLE-2024-BLG-0657, 
two K-type main-sequence stars for KMT-2024-BLG-2017, and a K-type and an early G-type 
main-sequence star for KMT-2024-BLG-2480.

We derived the angular Einstein radius using the relation $\thetae = \theta_*/\rho$, 
where $\theta_*$ is the angular radius of the source star and $\rho$ is the normalized 
source radius measured from the modeling. The angular source radius was inferred from 
the star's color and magnitude. To do this, we applied the empirical relation between 
the $(V-K, V)$ and $\theta_*$ provided by \citet{Kervella2004}. To use this relation, 
we first converted the measured $V - I$ color to $V - K$ using the color-color 
transformation from \citet{Bessell1988}, and then estimated $\theta_*$ based on the 
derived $V - K$ color.  With the estimated angular Einstein radius, we also computed 
the relative proper motion between the lens and source, given by $\mu = \thetae/\te$.

In Table~\ref{table:five}, we present the angular radii of the binary source stars, 
$\theta_{*,S_1}$ and $\theta_{*,S_2}$, along with the corresponding values of $\theta_{\rm E}$ 
and $\mu$. The angular Einstein radius $\thetae$ can be determined using either component 
of the binary source, that is, $\thetae = \theta_{*,S_1} / \rho_1$ or $\thetae = 
\theta_{*,S_2} / \rho_2$. In our analysis, we adopted the value derived from the 
better-constrained $\rho$ value. For the event KMT-2024-BLG-2480, the normalized source 
radius could not be determined because of the non-caustic-crossing nature of the anomaly. 
Consequently, values of $\theta_{\rm E}$ and $\mu$ are not provided for this event.

\section{Physical lens parameters} \label{sec:six}

In this section, we determine the physical parameters of the lens system, specifically 
the lens mass ($M$) and distance ($D_{\rm L}$). These parameters are constrained using 
the microlensing observables: the event timescale ($\te$), the angular Einstein radius 
($\thetae$), and the microlens parallax ($\pie$).  The observables are related to $M$ 
and $D_{\rm L}$ through the following equations:
\color{black}
\begin{equation}
\te = \frac{\thetae}{\mu},\qquad
\thetae = \sqrt{\kappa M \pirel}, \qquad
\pie = \frac{\pirel}{\thetae},
\label{eq5}
\end{equation}
where $\kappa = 4G/(c^2{\rm AU}) \simeq 8.14~\mathrm{mas}/M_\odot$, $\pirel = \mathrm{AU} 
\left(D_{\rm L}^{-1} - D_{\rm S}^{-1}\right)$ is the lens-source relative parallax, and 
$D_{\rm S}$ denotes the distance to the source \citep{Gould2000}.  Among these observables, 
the event timescale $\te$ is measured for all events in our sample. The angular Einstein 
radius $\thetae$ is measured for two events: OGLE-2024-BLG-0657 and KMT-2024-BLG-2017.  
However, the microlens parallax $\pie$ was not measured for any of the events.

To determine the physical parameters of the lens, we conducted a Bayesian analysis 
incorporating prior information based on the dynamical and spatial distributions of lens 
objects, as well as their mass distribution.  For the dynamical and spatial priors, we 
adopted the Galactic model of \citet{Jung2022}, and for the mass distribution, we employed 
the mass function proposed by \citet{Jung2021}. In the first stage of the analysis, we 
generated a large ensemble of synthetic microlensing events. Each simulated event was 
assigned a set of physical parameters, $(M_i, D_{{\rm L},i}, D_{{\rm S},i}, \mu_i)$, drawn 
from the adopted priors. In the second stage, we calculated the corresponding lensing 
observables, $\Omega_i = (t_{{\rm E},i}$, $\theta_{{\rm E},i})$, using the relations 
described in Eq.~(\ref{eq5}).  We then constructed posterior probability distributions 
for the lens mass and distance by assigning a statistical weight to each event:
\begin{equation}
w_i = \exp\left(-\frac{\chi_i^2}{2} \right); \qquad
\chi_i^2 = \sum_j \frac{(\Omega_{i,j} - \Omega_j)^2}{\sigma(\Omega_j)^2}.
\label{eq6}
\end{equation}
Here $\Omega_j$ represents the value of the lensing observable and $\sigma(\Omega_j)$ 
denotes its associated uncertainty. Finally, we inferred the posterior estimates of the 
physical parameters from these weighted distributions.

\begin{figure}[t]
\includegraphics[width=\columnwidth]{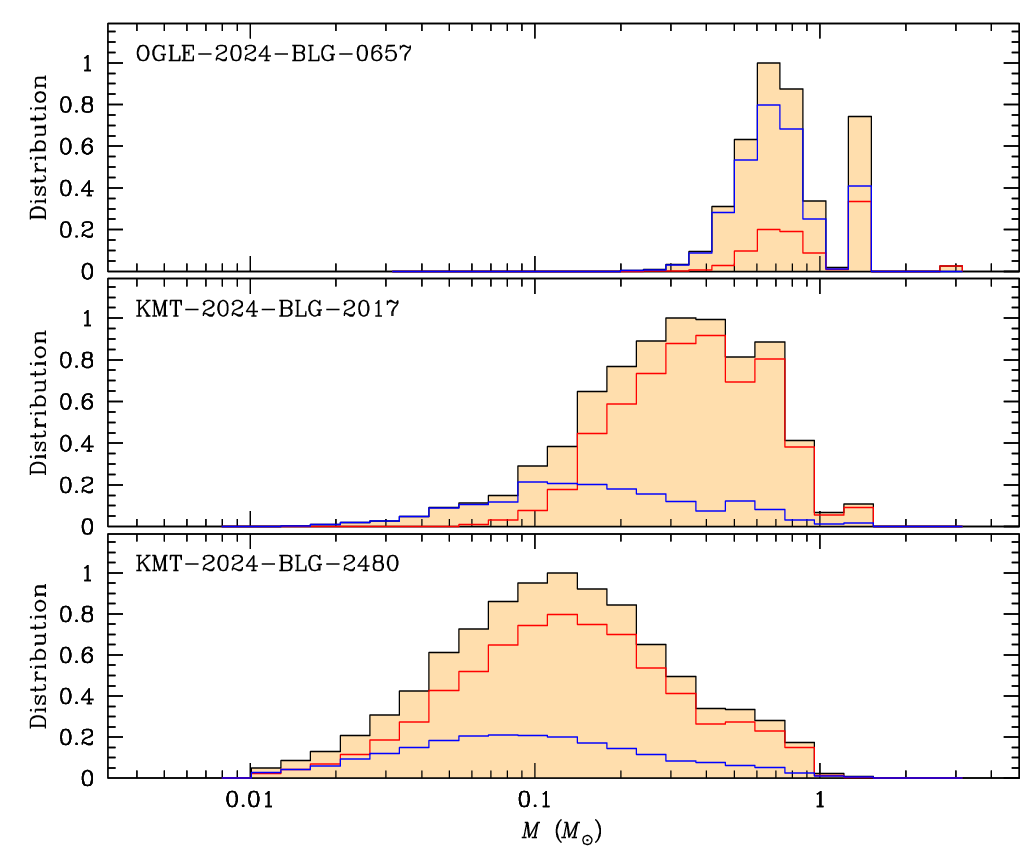}
\caption{
Bayesian posterior distributions of the primary lens mass ($M_1$) for the events. In 
each panel, the blue and red curves represent the posterior distributions arising 
from disk and bulge lens populations, respectively, while the black curve shows their 
combined distribution.
}
\label{fig:eight}
\end{figure}

In addition to the constraints from lensing observables, we imposed an additional constraint 
based on the blended flux. This arises from the fact that the lens contributes to the total 
blended light, and thus its flux cannot exceed the measured blend flux. For KMT-2024-BLG-2017, 
the blend is well characterized with $(V-I, I)_b \sim (1.4, 18.2)$. For OGLE-2024-BLG-0657, 
although the $V$-band magnitude is uncertain, the $I$-band blend flux is constrained to 
$I_b \sim 20.6$. In contrast, for KMT-2024-BLG-2480, neither the $I$- nor $V$-band 
blended magnitudes could be measured. For events with a measured $I$-band blend flux, we 
adopted $I_b$ as an upper limit on the lens brightness in the Bayesian analysis.

\begin{figure}[t]
\includegraphics[width=\columnwidth]{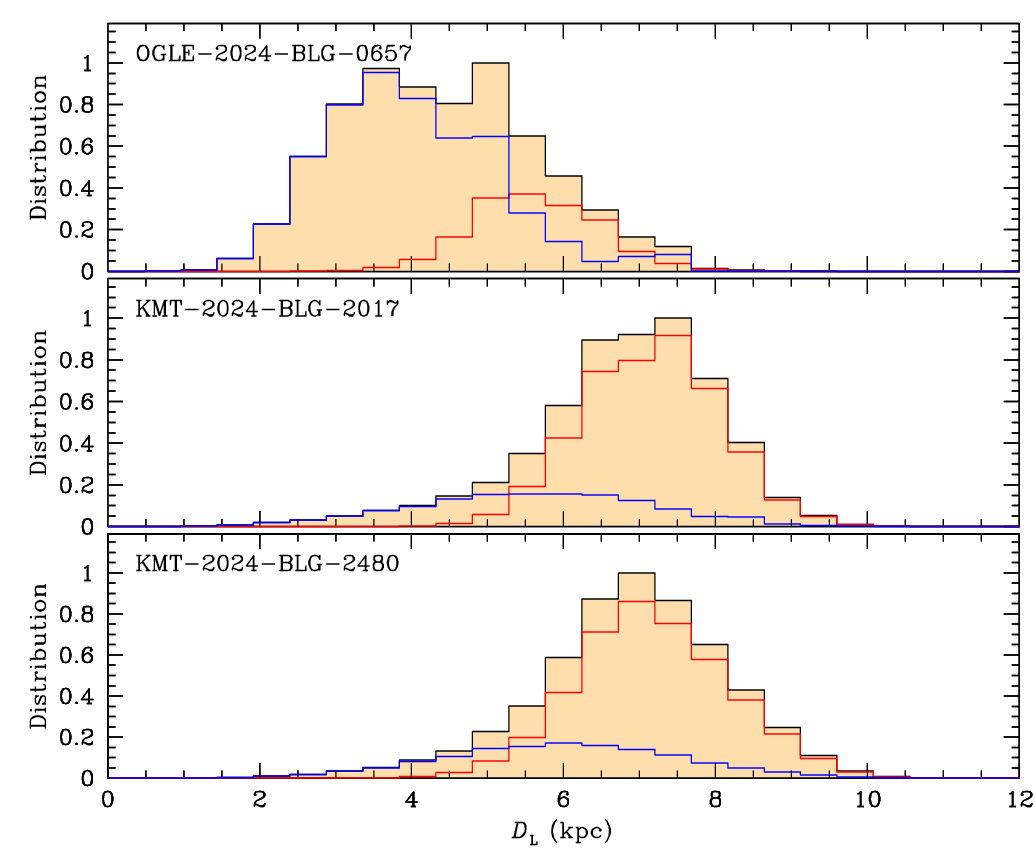}
\caption{
Bayesian posterior distributions of the lens distance. Notations are same as in 
Fig.~\ref{fig:eight}.
}
\label{fig:nine}
\end{figure}

The posterior distributions for lens mass and distance are presented in Figures~\ref{fig:eight} 
and ~\ref{fig:nine}, respectively, with the mass distributions corresponding to the primary 
lens component ($M_1$).  Based on the Bayesian posterior analysis, the lenses in KMT-2024-BLG-2017 
and KMT-2024-BLG-2480 are inferred to be binary systems composed of low-mass stars located in 
the Galactic bulge.  For KMT-2024-BLG-2017L, the binary lens components have masses
$M_1 = 0.36^{+0.35}_{-0.20}~M_\odot$ and
$M_2 = 0.22^{+0.20}_{-0.11}~M_\odot$,
with the system situated at a distance of
$\dl = 7.16^{+1.03}_{-1.29}$ kpc.
In the case of KMT-2024-BLG-2480, the close binary solution yields component masses of
$M_1 = 0.14^{+0.21}_{-0.08}~M_\odot$ and
$M_2 = 0.13^{+0.20}_{-0.08}~M_\odot$,
located at a distance of $\dl = 7.19^{+1.15}_{-1.22}$ kpc.
In both cases, the lensing objects are most likely situated within the Galactic bulge.
Alternatively, the wide binary solution suggests a significantly higher secondary lens mass 
of $M_2 \sim 3.0~M_\odot$.  This mass is physically implausible, not only because no such 
high-mass star exists in the bulge, but also because the flux from such a star would exceed 
the observed blended flux.

In contrast, the lens system OGLE-2024-BLG-0657L is likely a binary composed of two stellar 
remnants.  The Bayesian posterior distribution for $M_1$ reveals two distinct peaks at 
approximately 0.6--0.7~$M_\odot$ and 1.0--1.1~$M_\odot$, consistent with typical white 
dwarf and neutron star masses, respectively. Given its location at approximately 3.0 kpc, 
a luminous $M_1$ would have an apparent $I$-band magnitude of either $I_{\rm L} \sim 20.3$ 
or 17.7. In both cases, the lens would appear brighter than the observed blended flux of 
$I_b \sim 20.6$, which is inconsistent with the observations. Furthermore, the secondary 
lens component, being about 1.6 times more massive than the primary, would be even brighter 
if it were a main-sequence star.  These inconsistencies strongly suggest that both $M_1$ 
and $M_2$ are compact stellar remnants, a conclusion further supported by the stellar-type 
classifications in the Bayesian simulations.  The probabilities for the lens being in the 
Galactic disk and bulge are 76\% and 24\%, respectively, indicating that the lens is more 
likely to be located in the disk.

\section{Summary and conclusion} \label{sec:seven}

We conducted a comprehensive investigation of microlensing events detected by the OGLE 
and KMTNet surveys during the 2024 observing season, focusing particularly on events 
exhibiting complex anomaly features in their light curves. Through this investigation, 
we found that the light curves of three events OGLE-2024-BLG-0657, KMT-2024-BLG-2017, 
and KMT-2024-BLG-2480 exhibit anomaly features that cannot be adequately explained by 
the conventional three-body lensing models, such as the 2L1S or 1L2S configurations. 
These limitations highlighted the need for more advanced modeling approaches to properly 
interpret the observed data.

As a first step, we applied the 2L1S model to each light curve. While this model was 
able to account for the primary anomaly in each event, it proved insufficient to 
simultaneously explain the full set of complex features observed. This prompted us 
to explore more sophisticated four-body models by introducing an additional lens or 
source component.  Through this extended four-body modeling, we found that the full 
range of anomaly features in all three events could be successfully reproduced using 
a 2L2S configuration, in which both the lens and the source consist of two components.

We characterized the source systems by estimating the colors and magnitudes of the stars
comprising each binary source. This analysis revealed that the source in OGLE-2024-BLG-0657 
is a binary composed of G- and K-type main sequence stars. In the case of KMT-2024-BLG-2017, 
the source consists of two K-type main sequence stars. For KMT-2024-BLG-2480, the source 
was identified as a binary system consisting of a K-type star and an early G-type main 
sequence companion.

To estimate the physical parameters of the lens systems, we performed Bayesian analyses 
incorporating constraints from the measured microlensing observables and the blended 
flux. The results indicate that the lenses in KMT-2024-BLG-2017 and KMT-2024-BLG-2480 
are likely binary systems composed of low-mass stars located in the Galactic bulge, 
while the lens in OGLE-2024-BLG-0657L is inferred to be a binary system of stellar 
remnants such as white dwarfs or neutron stars residing in the Galactic disk. The 
remnant nature of the components in OGLE-2024-BLG-0657L may be confirmed through 
future high-resolution follow-up observations using adaptive optics on large-aperture 
telescopes. If either component is a luminous star, it will become detectable once it 
has sufficiently separated from the source. Conversely, if no lens flux is detected 
even after ample time has passed, this would strongly suggest that the lens is composed 
of dark remnants. These findings underscore the diversity of microlensing lens systems 
and highlight the importance of combining observational constraints with detailed 
modeling to reveal their underlying physical characteristics.

\begin{acknowledgments}
This research was supported by the Korea Astronomy and Space Science Institute under the R\&D 
program (Project No. 2025-1-830-05) supervised by the Ministry of Science and ICT.
This research has made use of the KMTNet system operated by the Korea Astronomy and Space Science 
Institute (KASI) at three host sites of CTIO in Chile, SAAO in South Africa, and SSO in Australia. 
Data transfer from the host site to KASI was supported by the Korea Research Environment Open NETwork 
(KREONET). 
C.Han acknowledge the support from the Korea Astronomy and Space Science Institute under the R\&D program 
(Project No. 2025-1-830-05) supervised by the Ministry of Science and ICT.
J.C.Y., I.G.S., and S.J.C. acknowledge support from NSF Grant No. AST-2108414. 
W.Zang acknowledges the support from the Harvard-Smithsonian Center for Astrophysics 
through the CfA Fellowship.
The OGLE project has received funding from the Polish National Science
Centre grant OPUS-28 2024/55/B/ST9/00447 to AU. 
\end{acknowledgments}



\bibliographystyle{aasjournal}
\bibliography{pasp_refs}

\end{document}